\documentclass[aps,prl,preprint,amsmath,amssymb,showpacs,preprintnumbers,superscriptaddress]{revtex4-1}
\usepackage{CJK}
\usepackage{graphicx}
\usepackage{dcolumn}
\usepackage{bm}
\usepackage{color}
\usepackage{hyperref}

\begin{document}
\graphicspath{{Figures/}}

\title{Toroidal states in $^{28}$Si with covariant density functional theory in 3D lattice space}
\author{Z. X. Ren}
\affiliation{State Key Laboratory of Nuclear Physics and Technology, School of Physics, Peking University, Beijing 100871, China}

\author{P. W. Zhao}
\affiliation{State Key Laboratory of Nuclear Physics and Technology, School of Physics, Peking University, Beijing 100871, China}

\author{S. Q. Zhang}
\affiliation{State Key Laboratory of Nuclear Physics and Technology, School of Physics, Peking University, Beijing 100871, China}

\author{J. Meng}
\email{mengj@pku.edu.cn}
\affiliation{State Key Laboratory of Nuclear Physics and Technology, School of Physics, Peking University, Beijing 100871, China}
\affiliation{Department of Physics, University of Stellenbosch, Stellenbosch, South Africa}
\affiliation{Yukawa Institute for Theoretical Physics, Kyoto University, Kyoto 606-8502, Japan}


\date{\today}

\begin{abstract}
  The toroidal states in $^{28}$Si with spin extending to extremely high are investigated with the cranking covariant density functional theory on a 3D lattice.
  Thirteen toroidal states with spin $I$ ranging from 0 to 56$\hbar$ are obtained, and their stabilities against particle emission are studied by analyzing the density distributions and potentials.
  The excitation energies of the toroidal states at $I=28$, 36, and 44$\hbar$ reasonably reproduce the observed three resonances extracted from the seven-$\alpha$ de-excitation of $^{28}$Si.
  The possible existence of $\alpha$ clustering in these toroidal states is discussed based on $\alpha$-localization function.
\end{abstract}

\maketitle

Most nuclei in their ground states are spherical or ellipsoidal~\cite{STONE2005deformation}.
Wheeler suggested that very heavy nuclei may have toroidal shapes due to the large Coulomb energies~\cite{Wheeler_toroidal}.
Pioneering works along this idea have been done by Wong~\cite{WONG1972Toroidal, Wong1973Troidal, Wong1978rotationg_torus}.
Based on the toroidal potential in radially displaced harmonic-oscillator model~\cite{WONG1972Toroidal, Wong1973Troidal}, it was found that although the toroidal states in liquid-drop model are unstable against sausage deformations, the nuclear shell effects may counterbalance with this instability~\cite{WONG1972Toroidal, Wong1973Troidal}.
Later on, Wong predicted that the toroidal states could be stabilized with a sufficiently high angular momentum by introducing an effective ``rotation'' about the symmetry axis to the toroidal states~\cite{Wong1978rotationg_torus}.
Recent investigation on shells in a toroidal nucleus in the intermediate-mass region with the toroidal potential can be found in Ref.~\cite{Wong2018Toroidal}.

The microscopic and self-consistent nuclear energy density functional theories (DFTs)~\cite{Bender2003Self, Vretenar2005PhysicsReport, meng2006PPNP, meng2016relativistic} have also been used to investigate the toroidal states in both  superheavy~\cite{Warda2007TOROIDAL, STASZCZAK2009heavy, Staszczak2017toroidal, AFANASJEV2018hyperheavy} and light nuclei~\cite{ZhangW2010Toroidal, Ichikaw2012Ca40_torus, STASZCZAK2014Toroidal, Staszczak2015TorusNneqZ, Ichikawa2014TDHF_Torus_short, Ichikawa2014TDHF_Torus_long}.
In particular, for the high-spin toroidal states, a toroidal state with an angular momentum of $60\hbar$ along the symmetry axis in $^{40}$Ca has been obtained with the cranking Skyrme DFT~\cite{Ichikaw2012Ca40_torus}.
Similar high-spin toroidal states in other nuclei with $28\leq A\leq52$ were also investigated in Refs.~\cite{STASZCZAK2014Toroidal, Staszczak2015TorusNneqZ, Ichikawa2014TDHF_Torus_short, Ichikawa2014TDHF_Torus_long}.

The angular momenta of the high-spin toroidal states are not from the collective rotation about the symmetry axis, but are generated by nucleon alignments~\cite{Ichikaw2012Ca40_torus, STASZCZAK2014Toroidal}.
The alignments of nucleons violate the time-reversal symmetries and, thus, induce strong currents~\cite{Ichikaw2012Ca40_torus},
which requires a proper treatment of the time-odd fields in the framework of DFTs.
In addition, some toroidal states may contain single particles in unbound states~\cite{Ichikaw2012Ca40_torus, Ichikawa2014TDHF_Torus_long, Staszczak2015TorusNneqZ}.
It is therefore important to examine the stability of the toroidal states against the nucleon emission.

To treat the time-odd fields and nucleon emission properly, a covariant DFT (CDFT) calculation in three-dimensional (3D) lattice space is preferred.
Due to the Lorentz invariance, the CDFT provides a self-consistent treatment of the time-odd fields; the time-odd fields share the same coupling constants as the time-even ones~\cite{Meng2013FT_TAC}.
Working in 3D lattice space makes it suitable to examine the nuclear stability against nucleon emission.
Fortunately, the CDFT in 3D lattice space is available now~\cite{tanimura20153d, REN2017Dirac3D, Ren2019C12LCS} after overcoming the longstanding problems of the variational collapse~\cite{ZhangIJMPE2010, hagino2010iterative} and Fermion doubling~\cite{tanimura20153d}.
In particular, the cranking CDFT in 3D lattice space is realized in Ref.~\cite{Ren2019C12LCS}, and thus provides a new opportunity to investigate the high-spin toroidal states.

By deep-inelastic collisions of $^{28}$Si on $^{12}$C target, an experiment has been performed recently to search for the high-spin toroidal states in $^{28}$Si~\cite{cao2019Toroidal}.
The excitation function for the seven-$\alpha$ de-excitation channel of $^{28}$Si reveals three resonances at the excitation energy region predicted in Ref.~\cite{STASZCZAK2014Toroidal}.
These three resonances are then suggested as the high-spin toroidal states in $^{28}$Si at spin $I=28$, $36$, and $44\hbar$, respectively.
The presence of these states is supported by the cranking CDFT calculations~\cite{cao2019Toroidal}.
In Ref.~\cite{cao2019Toroidal}, the discussions based on cranking CDFT calculations are mainly focused on the excitation energies.
It would be interesting to study the toroidal states in $^{28}$Si in details.
In this work, the toroidal states in $^{28}$Si will be investigated by the cranking CDFT in 3D lattice space systematically.


The starting point of the point-coupling density functional theory is a standard effective Lagrangian density of the form
\begin{equation}
  \mathcal{L}=\mathcal{L}^{\rm free}+\mathcal{L}^{\rm 4f}+\mathcal{L}^{\rm hot}+\mathcal{L}^{\rm der}+\mathcal{L}^{\rm em},
\end{equation}
including the Lagrangian density for free nucleons $\mathcal{L}^{\rm free}$,
the four-fermion point-coupling terms $\mathcal{L}^{\rm 4f}$, the higher order terms $\mathcal{L}^{\rm hot}$ accounting for the medium effects,
the derivative terms $\mathcal{L}^{\rm der}$ to simulate the finite-range effects that are crucial for a quantitative description of nuclear density distributions,
and the electromagnetic interaction terms $\mathcal{L}^{\rm em}$.
For the detailed formalism, one can refer to, for examples, Refs.~\cite{meng2016relativistic,NIKSIC2011PPNP,ZhaoPC-PK1}.
The nucleons in CDFT can also be coupled with finite-range meson fields, and the details can be found in Refs.~\cite{Long2004PK1, lalazissis2005new}.

For nuclear rotations, one can transform the effective Lagrangian into a rotating frame with a constant rotational frequency $\omega$ around the rotational axis~\cite{Zhao2011Ni60}.
By minimizing the Routhian of the total system, one can get the cranking CDFT~\cite{Afanasjev1999PhysicsReport, Meng2013FT_TAC, Zhao2018Spectroscopies}.
Assuming the rotational axis as the $z$-axis, the equation of motion for nucleon becomes,
\begin{equation}\label{Dirac_equation}
   \hat{h}'\psi_k=\left(\hat{h}_0-\omega\hat{j}_z\right)\psi_k=\varepsilon'_k\psi_k.
\end{equation}
Here $\hat{h}'$ is cranking Hamiltonian, $\varepsilon_k'$ is the single-particle Routhian, $-\omega\hat{j}_z$ is the Coriolis or cranking term, and $\hat{j}_z=\hat{l}_z+\frac{1}{2}\hat{\Sigma}_z$ is the $z$ component of the total angular momentum of the nucleon spinor.
From the single-particle wave functions $\psi_k$, the single-particle energies are obtained by calculating the expectation values of the single-particle Hamiltonian $\hat{h}_0$,
\begin{equation}\label{spHamiltonian}
   \hat{h}_0=\bm{\alpha\cdot}[-i\bm{\nabla}-\bm{V}(\bm{r})]+\beta[m_N+S(\bm{r})]+V_0(\bm{r}).
\end{equation}
The relativistic scalar $S(\bm{r})$ and vector $V_\mu(\bm{r})$ fields are connected in a self-consistent way to the densities and current distributions of the nucleons.
By solving the cranking Dirac equation \eqref{Dirac_equation} self-consistently, one can proceed to calculate various physical observables, such as angular momenta and total energies.
For the details, one can refer to, for examples, Refs.~\cite{Afanasjev1999PhysicsReport, Meng2013FT_TAC, meng2016relativistic, Zhao2018Spectroscopies}.

The cranking Dirac equations are usually solved in the harmonic oscillator basis~\cite{Koepf1989PAC_CDFT, Peng2008maganetic_roration, Zhao2011Ni60, Zhao2011PRL_AMR, Zhao2017ChiralRotation}.
In Ref.~\cite{Ren2019C12LCS}, the cranking CDFT is solved in 3D lattice space
by overcoming the variational collapse and the Fermion doubling problems with the inverse Hamiltonian~\cite{hagino2010iterative} and the Fourier spectral methods~\cite{Shen2011Spectral} respectively.

As noted in Ref.~\cite{Ren2019C12LCS}, when solving cranking Dirac equation in 3D lattice space,
the unphysical continuum with large angular momenta go down drastically and even cross the occupied single-particle Routhians at large rotational frequencies.
The occupations of these unphysical continuum would lead to an unphysical fission of the system.
To avoid this problem, as in Ref.~\cite{Ren2019C12LCS}, the cranking term $-\omega\hat{j}_z$ in Eq. \eqref{Dirac_equation} is replaced with a damped one $f_D(r)(-\omega\hat{j}_z)f_D(r)$, where $f_D(r)=\frac{1}{1+e^{(r-r_D)/a_D}}$ is a Fermi-type damping function with two parameters $r_D$ and $a_D$.

In the present cranking CDFT calculations for $^{28}$Si, the PC-PK1 density functional~\cite{ZhaoPC-PK1} is employed.
The 3D lattice is built in the Cartesian frame, and a step size 0.8 fm along the $x$, $y$, and $z$ axes is chosen.
The grid numbers are 34 for the $x$ and $y$ axes and 24 for the $z$ axis.
The size of the space adopted here is sufficient to obtain converged solutions.
The pairing correlations are neglected here, since they are substantially suppressed in the high-spin toroidal states.
The parameters in the Fermi-type damping function are $r_D=12$ fm and $a_D=0.2$ fm.



\begin{figure}[h!]
  \centering
  \includegraphics[width=0.9\textwidth]{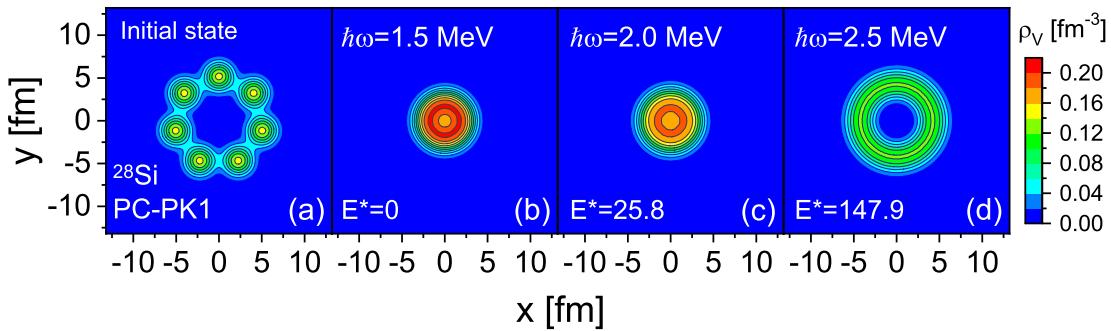}\\
  \caption{The total density distributions in the $z=0$ plane for (a) the initial state and
  (b)-(d) the states at rotational frequencies $\hbar\omega$=1.5, 2.0, and 2.5 MeV.
  The excitation energies with respect to the ground state are denoted by $E^*$ (in MeV).
  }\label{Fig1_InitialTorus}
\end{figure}
To obtain the toroidal states of $^{28}$Si, the initial state is constructed by placing seven wavefunction sets of $^{4}$He along a ring in the $z=0$ plane, as shown in Fig.~\ref{Fig1_InitialTorus}(a).
Self-consistent cranking CDFT calculations are then performed at different rotational frequencies with the rotational axis along the $z$-axis.
The density distributions for the finally converged states at $\hbar\omega=1.5$, 2.0, and 2.5 MeV are shown in Figs.~\ref{Fig1_InitialTorus}(b)-(d).
Fig.~\ref{Fig1_InitialTorus}(b) corresponds to the ground state of $^{28}$Si with the $z$-axis as the symmetry axis.
Figs.~\ref{Fig1_InitialTorus}(c) and (d) correspond to excited states with the excitation energies $E^*=25.8$ MeV and 147.9 MeV, respectively.
The density distribution shown in Fig.~\ref{Fig1_InitialTorus}(d) exhibits clearly a toroidal state.

For the toroidal state in Fig.~\ref{Fig1_InitialTorus}(d), the expectation value of total angular momentum $\hat{J}_z$ gives the spin $I=\langle J_z\rangle=44\hbar$, which is consistent with the results given by the Skyrme DFT and the toroidal potential~\cite{STASZCZAK2014Toroidal}.
Apart from the toroidal state at $I=44\hbar$, the results of the toroidal potential indicate that other toroidal configurations exist~\cite{STASZCZAK2014Toroidal}.

\begin{figure}[h!]
  \centering
  \includegraphics[width=0.45\textwidth]{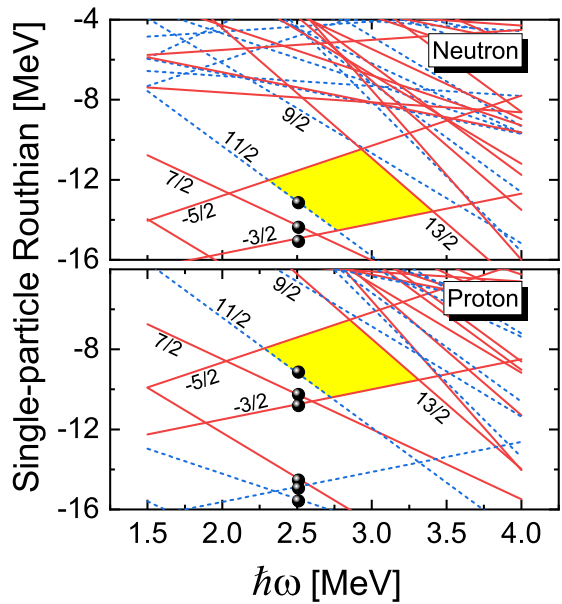}\\
  \caption{The single-particle Routhians as a function of the rotational frequency for the toroidal state with $I=44\hbar$.
  The solid and dashed lines represent the levels with positive and negative parities, respectively.
  The balls denote the occupied levels.
  The yellow areas represent the region where neutrons or protons occupy the lowest levels from the bottom of the potential.
  The $z$-components of total angular momenta for the corresponding single-particle levels are given.
  }\label{Fig2_SpRSpin44}
\end{figure}

In order to understand the toroidal configuration, the single-particle Routhians of the toroidal state at $I=44\hbar$ as a function of $\hbar\omega$ are shown in Fig.~\ref{Fig2_SpRSpin44}.
The occupied levels are denoted by balls.
The yellow areas represent the region of rotational frequency where neutrons or protons occupy the lowest levels from the bottom of the potential.
Particularly, in the region of rotational frequency 2.31$\sim$3.31 MeV, both proton and neutron occupy the lowest levels.
This indicates that the corresponding toroidal state is a local energy minimum with $I=44\hbar$.
All occupied levels are addressed as ``toroidal levels'', as their densities exhibit an axially-symmetric toroidal distribution.
As a result, the total density distribution, which is a composition of the density distributions of all occupied levels, is toroidal [see Fig.~\ref{Fig1_InitialTorus}(d)].

\begin{figure}[h!]
  \centering
  \includegraphics[width=0.45\textwidth]{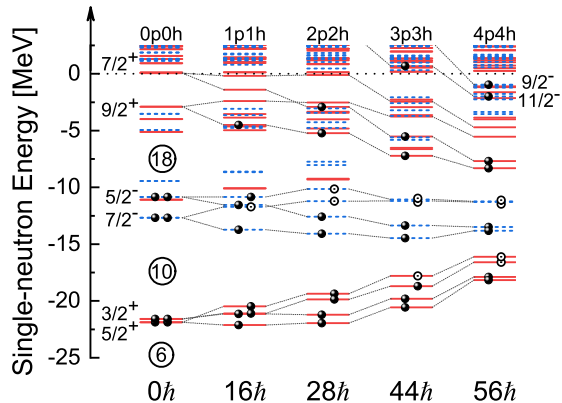}\\
  \caption{The single-neutron levels for the toroidal states with the symmetric proton and neutron configurations at $I=0$, 16, 28, 44, and 56$\hbar$.
  The solid and dashed lines represent the levels with positive- and negative-parity, respectively.
  The balls and circles denote the particles and holes in the toroidal levels (see text), respectively.
  The $z$-component of total angular momenta $m_z$ and parity $\pi$ are used to label the toroidal level as $|m_z|^\pi$.
  The balls and circles in the left (right) side denote the levels with positive (negative) $m_z$.
  For simplicity, the lowest six levels are not shown in the figure.
  }\label{Fig3_SpE}
\end{figure}

From the ``toroidal levels'', different toroidal configurations can be constructed.
Then the self-consistent calculations with fixed toroidal configurations thus obtained have been carried out in the framework of cranking CDFT.

For the symmetric proton and neutron configurations, the toroidal states at $I=0$, 16, 28, 44, and 56$\hbar$ have been found.
The corresponding single-neutron levels are shown in Fig.~\ref{Fig3_SpE}.
Due to the time-odd fields, the degeneracy of the single-neutron levels is lost for the toroidal states with $I\neq0$.
The toroidal states at $I=16$, 28, 44, and 56$\hbar$ respectively correspond to the [1p1h]$_{\nu,\pi}$, [2p2h]$_{\nu,\pi}$, [3p3h]$_{\nu,\pi}$, and [4p4h]$_{\nu,\pi}$ particle-hole excitations relative to the toroidal state at $I=0\hbar$.
These particle-hole excitations can only occur among the ``toroidal levels''.
The occupation of the other levels would contribute remarkable density in the center of the torus and destroy the toroidal structure.

\begin{figure}[h!]
  \centering
  \includegraphics[width=0.45\textwidth]{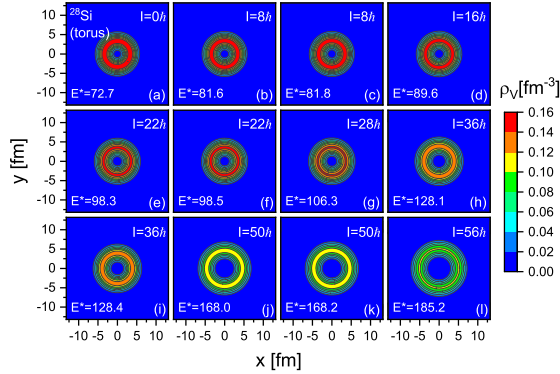}\\
  \caption{The total density distributions in the $z=0$ plane for the toroidal states obtained with fixed toroidal configurations.
  The spin $I$ ranges from 0$\hbar$ to 56$\hbar$.
  The excitation energies with respect to the ground state are denoted by $E^*$ (in MeV).
  }\label{Fig4_DensityPlane}
\end{figure}

For the asymmetric proton and neutron configurations, the toroidal states exist as well.
The toroidal configurations with [$n$p$n$h]$_\nu$[$(n+1)$p$(n+1)$h]$_\pi$ and [$(n+1)$p$(n+1)$h]$_\nu$[$n$p$n$h]$_\pi$ excitations are explored.
Their density distributions in the $z=0$ plane are shown in Fig.~\ref{Fig4_DensityPlane} together with the toroidal states with symmetric proton and neutron configurations.
Axially-symmetric toroidal distributions can be clearly seen here, though no symmetry is assumed a priori in the present 3D lattice cranking CDFT calculations.

In Figs.~\ref{Fig4_DensityPlane}(a)-(l), the spatial distributions of toroidal states become more diffuse with spin.
At $I=8$, 22, 36, and 50$\hbar$, the excitation energy of the toroidal state with [$n$p$n$h]$_\nu$[$(n+1)$p$(n+1)$h]$_\pi$ excitation is slightly lower than the one with [$(n+1)$p$(n+1)$h]$_\nu$[$n$p$n$h]$_\pi$ excitation.
This gentle difference mainly arises from the Coulomb energy, i.e.,
the Coulomb energy in the [$(n+1)$p$(n+1)$h]$_\pi$ excitation is slightly smaller than the one in [$n$p$n$h]$_\pi$ excitation.

The total density distributions shown in Fig.~\ref{Fig4_DensityPlane} can be well parameterized by a Gaussian distribution~\cite{Ichikaw2012Ca40_torus, STASZCZAK2014Toroidal, Ichikawa2014TDHF_Torus_long},
$\rho(x,y,z)=\rho_0e^{-[(\sqrt{x^2+y^2}-R_0)^2+z^2]/(d^2/\ln2)}$,
where $\rho_0$, $R_0$ and $d$ denote the maximum value of the nucleon density, the radius of the torus ring, and the width of a cross section of the torus ring, respectively.
It is worth mentioning that the width of the cross section of the torus ring is $d\sim1.3$ fm for all toroidal states here.
This value of $d$ is close to $d=1.3\sim1.4$ fm in Refs.~\cite{Ichikaw2012Ca40_torus, STASZCZAK2014Toroidal, Ichikawa2014TDHF_Torus_long}, as well as the width of an $\alpha$ particle $d_\alpha=1.46$ fm in Brink's $\alpha$-cluster model~\cite{Brink_alpha_cluster}.


As seen in Fig. \ref{Fig3_SpE}, for the toroidal state at $I=44\hbar$, the unbound neutron level with $m_z^\pi=11/2^-$ is occupied.
Similar occupations by neutrons and/or protons occur for other toroidal states as well, e.g., the ones at $I=22$, 28, 36, 50, and $56\hbar$.
The occupation of unbound levels raises the question whether the obtained toroidal states are stable against nucleon emission.

\begin{figure}[h!]
  \centering
  \includegraphics[width=0.45\textwidth]{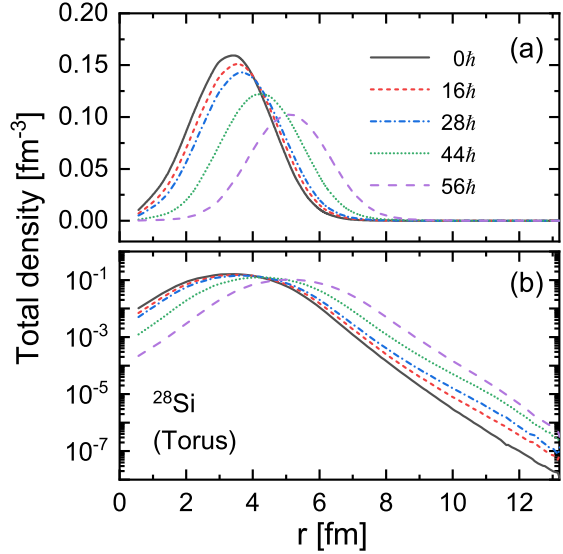}\\
  \caption{The total density distribution for toroidal states with the symmetric proton and neutron configurations at $I=0$, 16, 28, 44, and 56$\hbar$ in the $z=$0 plane as a function of the radial coordinate $r$ in normal (upper panel) and logarithmic scale (lower panel), respectively.
  }\label{Fig5_RadialDensity}
\end{figure}

\begin{figure}[h!]
  \centering
  \includegraphics[width=0.45\textwidth]{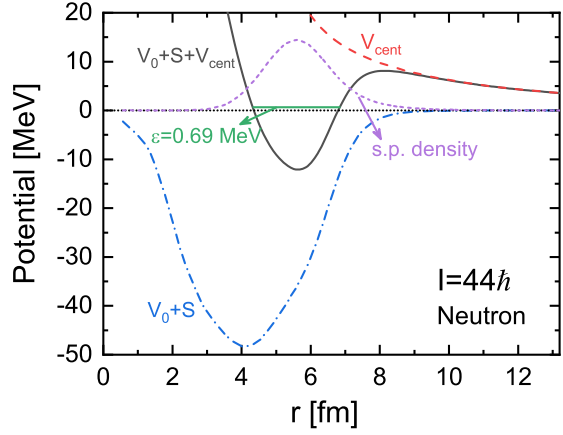}\\
  \caption{The radial distributions in the $z=0$ plane for the potential $V_0(r)+S(r)$, and the effective centrifugal potential $V_{\rm cent}(r)$ (see text), as well as their sum for the occupied unbound neutron level with $m_z^\pi=11/2^-$ for the toroidal state at $I=44\hbar$.
  The radial density profile multiplied with factor $r^2$ for this unbound level is shown.
  }\label{Fig6_RadialPotential}
\end{figure}

In Fig.~\ref{Fig5_RadialDensity}, the total density distributions for the toroidal states with the symmetric proton and neutron configurations at $I=0$, 16, 28, 44, and 56$\hbar$ in the $z=0$ plane are shown as a function of the radial coordinate $r$ in normal and logarithmic scale.
Due to the axial symmetry of the toroidal states, the density in the $z=0$ plane only depends on the radial coordinate, and decreases exponentially.
Therefore, one can conclude that they are localized.
This indicates that the toroidal states at $I=0$ and $16\hbar$ are stable against particle emission, while the ones at $I=28$, $44$, and $56\hbar$ are quasi-stable against particle emission due to the fact that the last occupied neutron or proton level has positive energy (see, e.g., Fig. \ref{Fig6_RadialPotential} for the $I=44\hbar$ state).
Similar conclusions hold true for the toroidal states with the asymmetric proton and neutron configurations.

In order to understand the localized density and the quasi-stable property of the unbound neutron level with $m_z^\pi=11/2^-$ against particle emission for the toroidal state at $I=44\hbar$,
the radial distributions in the $z=0$ plane for the potential $V_0(r)+S(r)$, and the effective centrifugal potential $V_{\rm cent}(r)$, as well as their sum are shown in Fig.~\ref{Fig6_RadialPotential}.
For the unbound neutron level, $m_z^\pi=11/2^-$ restricts its orbital angular momentum $l\geq5$.
Similar to the spherical Dirac equation~\cite{meng1998pseudospin, meng1998NPA},
an effective centrifugal potential,
\begin{equation}
  V_{\rm cent}(r)=\frac{1}{2M+S(r)-V_0(r)+\varepsilon}\frac{l(l+1)\hbar^2}{r^2},
\end{equation}
with $l=5$ and $\varepsilon$ the single-particle energy,
is plotted in Fig.~\ref{Fig6_RadialPotential} in the $z=0$ plane.
The centrifugal barrier is around 9 MeV and therefore the neutron level with $m_z^\pi=11/2^-$ is quasibound.
Accordingly, the radial density profile for this level is localized as shown in Fig. \ref{Fig6_RadialPotential}.
Similar analyses have been done for the other toroidal states, and
the corresponding density distributions are localized.

%
\begin{figure}[h!]
  \centering
  \includegraphics[width=0.45\textwidth]{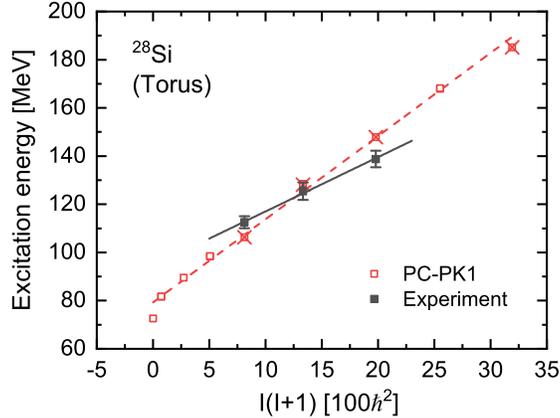}\\
  \caption{The excitation energies of toroidal states as a function of $I(I+1)$.
  The filled and open squares denote the observed and calculated energies, respectively.
  The lines represent the linear fitting with respect to $I(I+1)$.
  The open square with a cross denotes the toroidal state which is the local energy minimum at given spin.
  }\label{Fig7_ExciatationEnergy}
\end{figure}

In deep-inelastic collisions of $^{28}$Si on $^{12}$C target experiment~\cite{cao2019Toroidal}, three resonances with excitation energies 112.7, 125.4, and 138.7 MeV have been observed.
They are suggested as the toroidal states at spin $I=28$, 36, $44\hbar$, respectively.

In Fig.~\ref{Fig7_ExciatationEnergy}, the observed and calculated energies for the toroidal states are shown as a function of $I(I+1)$.
As discussed in Fig.~\ref{Fig2_SpRSpin44}, the toroidal state at $I=44\hbar$ is a local energy minimum.
Similar analyses have been done for the other toroidal states, and the toroidal states at $I=28$, $36$, $56\hbar$ are found to be local energy minima as well.
This is probably the reason for the observation of the three resonances, whose excitation energies are reasonably reproduced by the calculated ones at $I=28$, 36, $44\hbar$, respectively.
Although no sharp resonance corresponding to the toroidal state at $I=56\hbar$ is observed, its existence might be supported by the significant cross section observed above the resonance 138.7 MeV~\cite{cao2019Toroidal}.

The rotational band of the linear chain structure of seven-$\alpha$ clusters has also been calculated by performing the self-consistent and microscopic cranking CDFT calculations.
The highest excitation energy before the occurrence of fission is around 100 MeV.
This excitation energy is much smaller than the lowest observed resonance energy 112.7 MeV.
Therefore, the linear chain structure of seven-$\alpha$ clusters could be ruled out.

A linear relation between the excitation energies of the toroidal states and the $I(I+1)$ values is found for both the data and the calculated results.
This is consistent with the picture suggested by Bohr and Mottelson that the excitation associated with the alignment of single-particle levels along the symmetry axis possesses the similar behavior of a collective rotation~\cite{Bohr&Mottelson1975}.
The effective moments of inertia extracted from the observed and calculated energies are respectively 22.2 MeV$^{-1}\hbar^2$ and 14.5 MeV$^{-1}\hbar^2$.
There are several possible reasons for the difference of the experimental and theoretical moments of inertia.
One is that the excitation of toroidal states strongly depends on the single-particle levels, which are very challenging to be described accurately in a pure mean-field framework.
Another challenge is to take into account the beyond mean-field effects, such as the relative motion of individual $\alpha$ clusters (see similar discussions in Ref. \cite{Girod}), if the toroidal states are the systems of seven-$\alpha$ clusters.
The difference in the effective moment of inertia may also indicate the future prospects for probing the nuclear forces in the region of low nuclear density provided by the presence of toroidal light nuclei,
when toroidal states with other values of $I$ are experimentally located.

\begin{figure}[h!]
  \centering
  \includegraphics[width=0.45\textwidth]{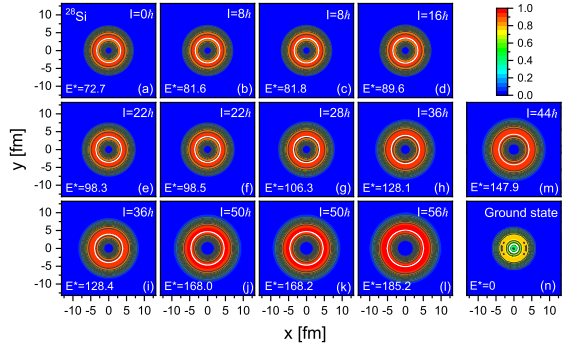}\\
  \caption{The distributions of the $\alpha$-localization function $\mathcal{C}_\alpha$ (see text) in the $z=0$ plane for the toroidal states with $I$ ranging from 0$\hbar$ to 56$\hbar$ [Figs.~(a)-(m)] and the ground state [Fig.~(n)].
  The maximum in density distribution for each toroidal state is denoted by white line.
  The excitation energies with respect to the ground state are denoted by $E^*$ (in MeV).
  }\label{Fig8_AlphaLocalization}
\end{figure}

The observed excitation function for the seven-$\alpha$ de-excitation channel of $^{28}$Si hints the $\alpha$-cluster structure in the corresponding resonance states.
It is interesting to investigate the $\alpha$-cluster structure in the toroidal states.
In Fig.~\ref{Fig4_DensityPlane}, it is already found that the width of the cross section of the torus density distributions is close to the width of an $\alpha$ particle used in Brink's $\alpha$-cluster model.
Further examination of the $\alpha$-cluster structure can be performed with the $\alpha$-localization function
which has been widely used to explore the $\alpha$-cluster structure~\cite{Reinhard2011Localization, ZhangCL2016NLF_fission, Schuetrumpf2017TDDFT_cluster, Ebran2017JPG_Cluster, Inakura2018Rod_shaped, tanimura2019clusterization}.
It is regarded as the first step to identifying $\alpha$ clustering, namely, the minimum necessary condition~\cite{Reinhard2011Localization}.

The $\alpha$-localization function is defined as $\mathcal{C}_\alpha=(\mathcal{C}_{n\uparrow}\mathcal{C}_{n\downarrow}\mathcal{C}_{p\uparrow}\mathcal{C}_{p\downarrow})^{1/4}$ with the nucleon localization function $\mathcal{C}_{q\sigma}(\bm{r})$~\cite{Reinhard2011Localization}.
A value of $\mathcal{C}_{q\sigma}(\bm{r})$ characterizes the probability of finding two nucleons with the same spin and isospin at neighborhood space.
For $\alpha$-cluster systems, $\mathcal{C}_{q\sigma}(\bm{r})\approx 1$,
and for homogeneous nuclear matter, $\mathcal{C}_{q\sigma}(\bm{r})\approx 1/2$.

In Fig.~\ref{Fig8_AlphaLocalization}, the distributions of $\mathcal{C}_\alpha$ in the $z=0$ plane for the toroidal states with $I$ ranging from 0 to 56$\hbar$ and the ground state are plotted.
The maximum in density distribution for each toroidal state and ground state is denoted by white line.
Around the white line, the $\mathcal{C}_\alpha$ values for the toroidal states are larger than 0.9, while those for the ground state are close to 0.5.
The feature of $\alpha$-localization function provides evidence of the possible existence of $\alpha$ clustering in the toroidal states.
It should be noted that different from the case of the linear chain structure of three-$\alpha$ clusters in $^{12}$C~\cite{Reinhard2011Localization}, the $\alpha$-localization functions and the density distributions of the toroidal states here do not show spatial separation due to the pure mean field picture, which is similar to many previous investigations~ \cite{ZhangW2010Toroidal, Ichikaw2012Ca40_torus, STASZCZAK2014Toroidal, Staszczak2015TorusNneqZ, Ichikawa2014TDHF_Torus_short, Ichikawa2014TDHF_Torus_long}.


In summary, the toroidal states in $^{28}$Si with spin extending to extremely high have been investigated with the cranking covariant density functional theory on a 3D lattice.
Thirteen toroidal states with the spin $I$ ranging from 0 to 56$\hbar$ are obtained.
The stabilities of these toroidal states against particle emission are illustrated by analyzing the density distributions and potentials.
The toroidal states at $I=28$, 36, 44$\hbar$ are local minima in energy at their given spins, and the corresponding excitation energies reasonably reproduce the observed three resonances extracted from the seven-$\alpha$ de-excitation of $^{28}$Si~\cite{cao2019Toroidal}.
A linear relation between the excitation energies of the toroidal states and the $I(I+1)$ values is found for both the data and the calculated results.
The $\alpha$-localization function $\mathcal{C}_\alpha$ has been calculated for each toroidal state, and the $\mathcal{C}_\alpha$ values are found to be larger than 0.9 around the maximum in density distribution.
This provides evidence of the possible existence of $\alpha$ clustering.

\begin{acknowledgments}
The authors thank C. Y. Wong for careful reading of the manuscript and valuable suggestions.
This work was partly supported by the National Key R\&D Program of China (Contracts No. 2018YFA0404400 and No. 2017YFE0116700) and
the National Natural Science Foundation of China (Grants No. 11621131001, No. 11875075, No. 11935003, and No. 11975031).
\end{acknowledgments}

%

\end{document}